\documentclass[%
 aip,
 amsmath,amssymb,
 reprint,%
]{revtex4-1}

\usepackage{graphicx}
\usepackage{dcolumn}
\usepackage{bm}

\usepackage[utf8]{inputenc}
\usepackage[T1]{fontenc}
\usepackage{mathptmx}
\usepackage{etoolbox}
\usepackage{tikz}
\usepackage{hyperref}
\usepackage{amsmath}
\usetikzlibrary{positioning,arrows.meta}
\usepackage{float}

\makeatletter
\def\@email#1#2{%
 \endgroup
 \patchcmd{\titleblock@produce}
  {\frontmatter@RRAPformat}
  {\frontmatter@RRAPformat{\produce@RRAP{*#1\href{mailto:#2}{#2}}}\frontmatter@RRAPformat}
  {}{}
}%
\makeatother
\begin{document}

\preprint{AIP/123-QED}

\title[i-Rheo-Tempo]{i-Rheo-Tempo: A Model-Free, Quadrature-Free Reconstruction of the Shear Relaxation Modulus from Complex Viscosity}

\author{Jorge Ram\'irez}
 \email{jorge.ramirez@upm.es}
\affiliation{ 
Departamento de Ingenier\'ia Qu\'imica, Universidad Polit\'ecnica de Madrid, Jos\'e Guti\'errez Abascal 2, 28006 Madrid, Spain
}%

\author{Manlio Tassieri}
 \email{manlio.tassieri@glasgow.ac.uk}
\affiliation{%
Division of Biomedical Engineering, James Watt School of Engineering, Advanced Research Centre, University of Glasgow, Glasgow, G11 6EW, UK
}%

\date{\today}

\begin{abstract}
Reliable transformation between frequency- and time-domain material functions remains a central challenge in linear viscoelasticity due to finite bandwidth, discrete sampling, and experimental noise.
We introduce \emph{i\text{-}Rheo-Tempo}, a quadrature-free method that reconstructs the shear relaxation modulus directly from dynamic measurements through an exact second-derivative representation of the complex viscosity. When the spectrum is approximated as piecewise linear, the inversion reduces to a compact interval-slope formulation based solely on local spectral properties, avoiding numerical quadrature, parametric fitting, and predefined relaxation spectra.
The method is validated against a set of complex fluids including synthetic models, polymer melts, industrial elastomers, comb polymers, and broadband microrheology datasets spanning nearly nine decades in frequency. In all cases, the reconstructed relaxation modulus is in quantitative agreement with independent time-domain measurements.
These results demonstrate that \emph{i\text{-}Rheo-Tempo} provides a robust, model-free solution to the frequency-to-time inverse problem and, more generally, establishes a framework for recovering time-domain responses from experimentally measured complex spectra.
\end{abstract}

\maketitle

\section{\label{sec:level1}Introduction}

"\textit{One of the problems frequently occurring in the investigation of the relaxation behaviour of linear viscoelastic materials is that of converting results of dynamic measurements into the result of transient
experiments, viz. creep or stress relaxation, or vice versa.}"\cite{schwarzl1975}

This observation, articulated by Schwarzl in 1973, succinctly captures what remains one of the central challenges of linear viscoelasticity: the reliable transformation between frequency-domain measurements and time-domain material functions.
Formally, the relationship between the frequency-dependent dynamic moduli and the time-dependent shear relaxation modulus is defined exactly through Fourier transformation.\cite{Ferry1980}
The mathematical problem is therefore well posed. In practice, however, experimental spectra are finite in bandwidth, discretely sampled, and inevitably affected by measurement noise. Direct numerical evaluation of Fourier integrals—whether in the forward or inverse direction—under such conditions is well known to generate oscillations, truncation artifacts, and strong sensitivity to endpoint treatment.

Since Schwarzl’s seminal contributions, numerous numerical procedures have been proposed to approximate these transformations in both directions. These include quadrature-based Fourier inversions, recursive schemes, Laplace-transform approaches with analytic continuation, and discrete approximation formulas\cite{schwarzl1968, baumgaertelDeterminationDiscreteRelaxation1989, kamathDeterminationPolymerRelaxation1989a, honerkampNonlinearRegularizationMethod1993,  takehComputerProgramExtract2013, shanbhagRelaxationSpectraUsing2020}. While mathematically rigorous, their practical implementation remains sensitive to spectral truncation, interpolation strategy, and boundary completion.
As a consequence, the generalized Maxwell model, or similar discrete spectral constructions,\cite{helfer2025} has become the default operational framework for interconverting time- and frequency-domain data. By representing the material response as a finite weighted sum of relaxation modes, one obtains an analytic form that can be transformed exactly between time and frequency domains. While computationally robust and convenient, this procedure is inherently model dependent and sensitive to the chosen number of modes\cite{tassieri2018}: the reconstructed material function reflects the imposed spectral structure as much as the experimental observations.

An alternative approach was introduced by Evans \textit{et al}.\cite{evans2009} and further developed by Tassieri \textit{et al}.\cite{tassieri2016} within the i-Rheo framework, where Fourier transformations were reformulated through a second-derivative representation of the measured response functions in the sense of distributions. In this formulation, numerical quadrature is replaced by exact evaluations of slope discontinuities (slope jumps), substantially reducing discretization artifacts while preserving full model independence.

In this work, we extend the derivative‑based frequency–to–time reconstruction to the complex‑viscosity domain. We derive an exact second‑derivative formulation of the inverse Fourier transform and implement it directly on experimental frequency nodes, yielding a closed‑form delta‑sum expression for the shear relaxation modulus. This framework, termed \emph{i\text{-}Rheo-Tempo}, eliminates numerical quadrature and avoids imposing predefined relaxation spectra or parametric models. Its accuracy and robustness are assessed across progressively more demanding test cases encompassing diverse molecular architectures and measurement bandwidths.

We first validate the method using synthetic data from analytically tractactable models, notably the Burgers model, whose multiple relaxation modes and curved spectrum provide a stringent benchmark independent of experimental uncertainty.

We then reanalyse datasets previously examined with the original i‑Rheo approach~\cite{tassieri2016}, including: an industrial styrene–butadiene rubber, and a monodisperse linear polyisoprene melt.
For each system, independently measured step‑strain data provide the reference relaxation modulus $G(t)$. Comparing these measurements with \emph{i\text{-}Rheo-Tempo} reconstructions from dynamic data verifies consistency between the time‑ and frequency‑domain derivative formulations.

As additional representative test case, we consider well-entangled linear polymer melts inspired by Rubinstein and Colby~\cite{rubinstein1988, rubinstein2003}, comprising highly monodisperse polybutadiene samples with different molecular weights. These highly monodisperse systems exhibit the characteristic features of entangled polymer dynamics, including a well-defined rubbery plateau and a sharp transition to the terminal regime, thereby providing a stringent benchmark for assessing the performance of the method.

We further analyse model comb polymers from Kapnistos and Vlassopoulos~\cite{kapnistos2009}. Their nearly monodisperse branched architectures generate hierarchical relaxation due to branch retraction and backbone reptation, yielding broad multi‑step stress‑relaxation profiles. Because both dynamic moduli and independent stress-relaxation measurements are available, spanning ten decades in frequency and time, these systems provide a fully experimental validation of the second-derivative discretization across multiple relaxation mechanisms.

Finally, we apply the method to broadband diffusing-wave spectroscopy microrheology data reported by Scheffold and co-workers~\cite{Scheffold2026}, which span nine decades in frequency. These measurements probe high-frequency, inertia-affected regimes and thus represent the most stringent test in terms of bandwidth and sensitivity to boundary artefacts. In this context, \emph{i\text{-}Rheo-Tempo} opens new opportunities for extracting time-domain information with unprecedented resolution from experimentally accessible spectra, enabling access to dynamical regimes that were previously difficult to quantify reliably.

Together, these cases—ranging from monodisperse entangled melts and industrial elastomers to architecturally complex comb polymers and ultra‑broadband microrheology—demonstrate that \emph{i\text{-}Rheo-Tempo} provides a quadrature‑free, model‑independent, and physically consistent solution to the frequency‑to‑time viscoelastic inverse problem directly compatible with experimental dynamic measurements.

\section{\label{sec:level2}THEORETICAL BACKGROUND}

\subsection{Fourier Convention and One-Sided Representation}

Throughout this work we adopt the physics Fourier-transform convention
\begin{equation}
\hat f(\omega)
=
\int_{-\infty}^{+\infty}
f(t)\,e^{-i\omega t}\,dt,
\qquad
f(t)
=
\frac{1}{2\pi}
\int_{-\infty}^{+\infty}
\hat f(\omega)\,e^{i\omega t}\,d\omega,
\label{eq:fourier_pair}
\end{equation}
where $t\in\mathbb{R}$ denotes time, $\omega\in\mathbb{R}$ the angular frequency, $f(t)$ is assumed to be integrable (or square-integrable) over $\mathbb{R}$, and $\hat f(\omega)$ denotes its Fourier transform.

In linear viscoelasticity, the material functions considered are real-valued causal functions of time, i.e.\ $f(t)=0$ for $t<0$. For any real time-domain signal $f(t)\in\mathbb{R}$, its Fourier transform satisfies Hermitian symmetry
\begin{equation}
\hat f(-\omega)=\hat f^*(\omega),
\end{equation}
where $^*$ denotes complex conjugation.

Writing
\begin{equation}
\hat f(\omega)=R(\omega)+iI(\omega),
\end{equation}
with $R(\omega),I(\omega)\in\mathbb{R}$ representing the real and imaginary
parts of the spectrum, the inverse transform Eq.~\eqref{eq:fourier_pair} can be expressed using positive frequencies only.
Exploiting Hermitian symmetry yields the one-sided representation
\begin{equation}
f(t)
=
\frac{1}{\pi}
\int_{0}^{+\infty}
\left[
R(\omega)\cos(\omega t)
-
I(\omega)\sin(\omega t)
\right]
\,d\omega,
\qquad t\in\mathbb{R}.
\label{eq:onesided_general}
\end{equation}

Equation~\eqref{eq:onesided_general} forms the starting point for the derivative-based reformulation developed in the following sections.

\subsection{Frequency-to-Time Relations for the Shear Relaxation Modulus}

The relation between frequency-domain measurements and the time-domain relaxation modulus is embodied in the classical integral expressions reported by Ferry (Chapter~3, Eqs.~41--42)~\cite{Ferry1980}, which we recall here for convenience:
\begin{align}
G(t)
&=
G_e
+
\frac{2}{\pi}
\int_{0}^{+\infty}
\frac{G''(\omega)}{\omega}
\cos(\omega t)\,d\omega,
\label{eq:ferry_general_sin}
\\
G(t)
&=
G_e
+
\frac{2}{\pi}
\int_{0}^{+\infty}
\frac{G'(\omega)-G_e}{\omega}
\sin(\omega t)\,d\omega,
\label{eq:ferry_general_cos}
\end{align}
for $t>0$, provided the integrals converge. Here, $G(t)$ denotes the shear relaxation modulus, $G'(\omega)$ and $G''(\omega)$ are the storage and loss moduli, $\omega$ is the angular frequency, and $G_e = \lim_{t\to\infty} G(t)$ is the equilibrium modulus, representing any residual elastic response at long times.

Equations~\eqref{eq:ferry_general_sin} and \eqref{eq:ferry_general_cos} are mathematically equivalent and apply to both viscoelastic solids ($G_e>0$) and viscoelastic fluids ($G_e=0$). The subtraction of $G_e$ in the cosine branch ensures convergence of the integral when a finite equilibrium modulus is present.

Despite their exactness, direct numerical evaluation of these relations from experimental data is well known to be ill-conditioned due to finite bandwidth, discrete sampling, and sensitivity to boundary treatment. \textit{It is precisely at this point that the present formulation departs from conventional approaches.}

For complex fluids ($G_e=0$), the above relations reduce to
\begin{align}
G(t)
&=
\frac{2}{\pi}
\int_{0}^{+\infty}
\frac{G''(\omega)}{\omega}
\cos(\omega t)\,d\omega,
\label{eq:ferry_sin}
\\
G(t)
&=
\frac{2}{\pi}
\int_{0}^{+\infty}
\frac{G'(\omega)}{\omega}
\sin(\omega t)\,d\omega.
\label{eq:ferry_cos}
\end{align}

Under the same Fourier convention, the complex viscosity is defined as
\begin{equation}
\eta^*(\omega)
=
\frac{G^*(\omega)}{i\omega}
=
\eta'(\omega)-i\eta''(\omega),
\qquad \omega>0,
\end{equation}
with
\begin{equation}
\eta'(\omega)
=
\frac{G''(\omega)}{\omega},
\qquad
\eta''(\omega)
=
\frac{G'(\omega)}{\omega}.
\end{equation}

Substitution into Eqs.~\eqref{eq:ferry_sin}--\eqref{eq:ferry_cos} yields the equivalent viscosity-based forms
\begin{align}
G(t)
&=
\frac{2}{\pi}
\int_{0}^{+\infty}
\eta'(\omega)
\cos(\omega t)\,d\omega,
\label{eq:G_eta2_base}
\\
G(t)
&=
\frac{2}{\pi}
\int_{0}^{+\infty}
\eta''(\omega)
\sin(\omega t)\,d\omega.
\label{eq:G_eta1_base}
\end{align}

In the present work we restrict attention to complex fluids, for which $G_e=0$, and Eqs.~\eqref{eq:G_eta2_base} and \eqref{eq:G_eta1_base} constitute the starting point of the \emph{i\text{-}Rheo-Tempo} framework.

\begin{figure}[!t]
\centering
\begin{tikzpicture}[
    node distance=7mm,
    >=stealth,
    every node/.style={font=\footnotesize},
    block/.style={
        rectangle,
        draw,
        rounded corners,
        text width=42mm,
        minimum height=8mm,
        align=center,
        thick,
        inner sep=3pt
    }
]

\node[block] (data)
    {Experimental data\\
    $\{\omega_k,\,G'(\omega_k),\,G''(\omega_k)\}$};

\node[block, below=of data] (eta)
    {Compute $\eta^*(\omega)=G^*(\omega)/(i\omega)$\\
     and extract $\eta'(\omega)$, $\eta''(\omega)$};

\node[block, below=of eta] (lfbc)
    {Low-frequency conditioning\\
     local fit for $\eta'(0)$\\
     and imposed $\eta''(0)$};

\node[block, below=of lfbc] (hfext)
    {High-frequency completion\\
     local polynomial fit near $\omega_{\max}$\\
     and limited spectral extension};

\node[block, below=of hfext] (interp)
    {Interpolation in linear $\omega$\\
     and logarithmic resampling};

\node[block, below=of interp] (piecewise)
    {Piecewise-linear spectrum\\
     with interval slopes $a_k$};

\node[block, below=of piecewise] (inv)
    {Interval-slope inversion\\
     $G(t)=\dfrac{2}{\pi t^2}\sum_k a_k\,[K_k(t)-K_{k+1}(t)]$\\
     $+\,(2/\pi)\,\eta''(0)/t$};

\node[block, below=of inv] (thr)
    {Long-time robustness criterion\\
     threshold estimated from the first\\
     low-frequency interval};

\node[block, below=of thr] (tail)
    {Optional terminal reference\\
     single-Maxwell fit from the\\
     low-frequency crossover of the moduli};

\node[block, below=of tail] (final)
    {Final relaxation modulus\\
     reported on
     $\left[1/\omega_{\max},\,1/\omega_{\min}\right]$};

\draw[->] (data) -- (eta);
\draw[->] (eta) -- (lfbc);
\draw[->] (lfbc) -- (hfext);
\draw[->] (hfext) -- (interp);
\draw[->] (interp) -- (piecewise);
\draw[->] (piecewise) -- (inv);
\draw[->] (inv) -- (thr);
\draw[->] (thr) -- (tail);
\draw[->] (tail) -- (final);

\end{tikzpicture}
\caption{\textbf{Workflow of the \emph{i\text{-}Rheo-Tempo} method.} Experimental dynamic moduli are converted into complex viscosity, conditioned at low and high frequency, interpolated and resampled, and then inverted through the interval-slope formulation to reconstruct the relaxation modulus. A robustness threshold is estimated from the first low-frequency interval to identify the onset of the unreliable long-time tail. Optionally, a single-Maxwell fit based on the low-frequency crossover of the moduli may be superimposed as a physically constrained terminal reference.}
\label{fig:workflow}
\end{figure}
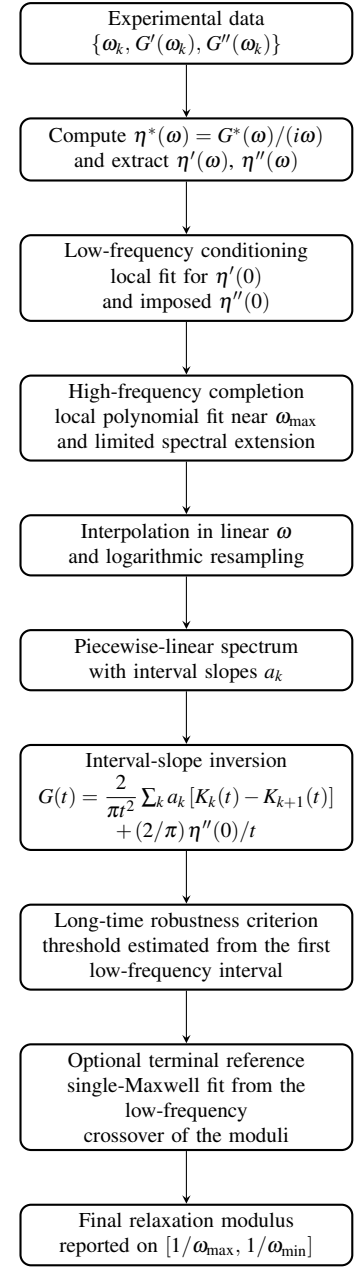

\section{Methods}
The overall workflow of \emph{i\text{-}Rheo-Tempo} is summarised in Fig.~\ref{fig:workflow}. The method transforms experimental frequency-domain data into a time-domain relaxation modulus through a sequence of well-defined steps—conditioning, interpolation, and interval-slope inversion—each designed to preserve physical consistency while avoiding numerical quadrature. The individual components of this workflow are elucidated in detail in the sections that follow.

\subsection{Second-Derivative Representation of the Inverse Transform}

Starting from the viscosity-based relations Eqs.~\eqref{eq:G_eta2_base} and \eqref{eq:G_eta1_base},
the inverse transform can be reformulated through two successive integrations by parts with respect to the angular frequency $\omega$.
To make the derivation transparent, we briefly outline the key steps. This procedure transfers the frequency dependence from the harmonic kernel to the curvature of the spectrum. Under standard physical conditions, the boundary terms vanish, yielding a representation in terms of the second derivative of the complex viscosity.

Let us consider first the sine branch.
Assuming sufficient smoothness of $\eta''(\omega)$, boundedness at $\omega=0$, and decay of $\eta''(\omega)$ and its first derivative as $\omega\to\infty$, two integrations by parts yield
\begin{equation}
G(t)
=
\frac{2}{\pi}
\left[
\frac{\eta''(0)}{t}
-
\frac{1}{t^{2}}
\int_{0}^{+\infty}
\ddot{\eta}''(\omega)
\sin(\omega t)\,d\omega
\right].
\label{eq:G_eta2_point4}
\end{equation}
where $\ddot{\eta}''(\omega)$ referes to the second derivative of $\eta''(\omega)$ with respect to the frequency. All boundary contributions at $\omega\to\infty$ vanish under the assumed decay conditions, leaving only the endpoint term $\eta''(0)/t$.
For viscoelastic fluids, the low-frequency condition $\eta''(0)=0$ holds identically, so that Eq.~\eqref{eq:G_eta2_point4} reduces to
\begin{equation}
G(t)
=
-
\frac{2}{\pi t^{2}}
\int_{0}^{+\infty}
\ddot{\eta}''(\omega)
\sin(\omega t)\,d\omega.
\label{eq:G_eta2_point4_fluid}
\end{equation}

Proceeding analogously for the cosine branch gives
\begin{equation}
G(t)
=
-
\frac{2}{\pi t^{2}}
\int_{0}^{+\infty}
\ddot{\eta}'(\omega)
\cos(\omega t)\,d\omega,
\label{eq:G_eta1_point4}
\end{equation}
provided $\eta'(\omega)$ and its first derivative remain bounded at the spectral limits.

Equations~\eqref{eq:G_eta2_point4_fluid} and \eqref{eq:G_eta1_point4}
constitute the continuous second-derivative representation underlying \emph{i\text{-}Rheo-Tempo}. Their essential feature is that the inverse transform depends not on the spectral functions themselves, but on their curvature with respect to frequency. This reformulation is the basis of the model-free discretisation described below.

\subsection{Distributional and Interval-Slope Discretisation}

Experimental spectra are available only at discrete angular frequencies $\{\omega_k\}_{k=1}^{N}$, with no information between sampling nodes.
To construct a mathematically consistent inverse transform without introducing preconceived relaxation spectra, the viscosity components must therefore be represented in a form compatible with differentiation in the distributional sense.

Let $f(\omega)$ denote a real-valued spectral function sampled at the nodes $\{\omega_k\}_{k=1}^{N}$ and approximated as piecewise linear between consecutive frequencies:
\begin{equation}
f(\omega)=a_k\,\omega+b_k,
\qquad
\omega\in[\omega_k,\omega_{k+1}],
\qquad
k=1,\dots,N-1,
\end{equation}
with interval slopes
\begin{equation}
a_k=
\frac{f(\omega_{k+1})-f(\omega_k)}
{\omega_{k+1}-\omega_k}.
\label{eq:interval_slopes_generic}
\end{equation}

Under this representation, the first derivative is piecewise constant, whereas the second derivative vanishes everywhere except at the nodes where the slope changes. In the sense of distributions,
\begin{equation}
f''(\omega)
=
\sum_{k=2}^{N-1}
\Delta a_k\,\delta(\omega-\omega_k),
\qquad
\Delta a_k=a_k-a_{k-1},
\label{eq:distributional_second_derivative}
\end{equation}
where $\delta(\omega-\omega_k)$ denotes the Dirac delta distribution.
Equation~\eqref{eq:distributional_second_derivative}
expresses the fact that spectral curvature is concentrated entirely at the slope discontinuities.

Applied directly to the viscosity components, this gives the interior
distributional representation
\begin{align}
G_{\eta''}^{\mathrm{int}}(t)
&=
-
\frac{2}{\pi t^2}
\sum_{k=2}^{N-1}
\Delta a_k^{(2)}
\sin(\omega_k t),
\label{eq:G_eta2_deltasum}
\\
G_{\eta'}^{\mathrm{int}}(t)
&=
-
\frac{2}{\pi t^2}
\sum_{k=2}^{N-1}
\Delta a_k^{(1)}
\cos(\omega_k t),
\label{eq:G_eta1_deltasum}
\end{align}

where $a_k^{(1)}$ and $a_k^{(2)}$ are the local slopes of $\eta'(\omega)$ and $\eta''(\omega)$, respectively, evaluated over each interval $[\omega_k,\omega_{k+1}]$.
These expressions are exact for the piecewise-linear spectrum on the interior interval $[\omega_1,\omega_N]$. The summations start at $k=2$ and end at $k=N-1$ because the first and last experimental nodes do not correspond to interior curvature terms; rather, they generate the boundary-slope contributions arising explicitly from integration by parts.

The numerical formulation used in \emph{i\text{-}Rheo-Tempo} is obtained by rewriting the interior jump representation in terms of the interval slopes themselves. Combining the interior distributional contribution with the associated boundary terms yields an interval-sum form in which each interval contributes through the difference of neighbouring harmonic kernels. Thus, for the real part of the complex viscosity,
\begin{equation}\begin{split}
&\frac{a_{N}^{(1)}\cos(\omega_N t)-a_1^{(1)}\cos(\omega_1 t)}{t^2} - \frac{1}{t^2}
\sum_{k=2}^{N-1}\Delta a_k^{(1)}\cos(\omega_k t)
=\\
&\frac{1}{t^2}\sum_{k=1}^{N-1}a_k^{(1)}\bigl[\cos(\omega_{k+1} t)-\cos(\omega_{k} t)\bigr],
\end{split}
\label{eq:cosine_telescoping}
\end{equation}
whereas for the imaginary part,
\begin{equation}\begin{split}
&\frac{a_{N}^{(2)}\sin(\omega_N t)-a_1^{(2)}\sin(\omega_1 t)}{t^2}
-
\frac{1}{t^2}
\sum_{k=2}^{N-1}
\Delta a_k^{(2)}\sin(\omega_k t)
=\\
&\frac{1}{t^2}
\sum_{k=1}^{N-1}
a_k^{(2)}
\bigl[
\sin(\omega_{k+1} t)-\sin(\omega_k t)
\bigr].
\end{split}
\label{eq:sine_telescoping}
\end{equation}

These relations are algebraically equivalent to the distributional representation, but they are numerically preferable because they express the reconstruction directly as a sum of interval contributions. In this form, the expressions exhibit a structure where each slope multiplies the difference between neighbouring harmonic kernels, so that adjacent contributions \textit{partially} cancel and only variations in the local slopes are retained. As a result, the reconstruction is naturally governed by the slope jumps, which encode the local curvature of the spectrum, while boundary slopes are incorporated consistently without introducing artificial endpoint singularities associated with zero-slope continuation.

Notably, these relations are of general validity for the inverse Fourier transform of any experimentally sampled complex function whose real and imaginary components satisfy the standard integrability conditions: (i) they remain finite at $\omega=0$, and (ii) they vanish in the limit $\omega\to\infty$.

\subsection{Boundary Conditioning and Spectral Completion}

The second-derivative formulation developed above is exact only when the complex-viscosity spectrum is defined over the full semi-infinite domain $\omega\in[0,\infty[$. Experimental data, however, occupy a finite and necessarily truncated window $[\omega_{\min}^{\mathrm{exp}},\,\omega_{\max}^{\mathrm{exp}}]$.
Outside this interval no information is available, whereas the inverse transform remains intrinsically sensitive to both low- and high-frequency behaviour.

The practical difficulty is particularly severe at low frequencies. In the original endpoint-jump formulation, truncation of the measured spectrum implicitly enforced an artificial zero-slope continuation outside the experimental window. This generated unphysical boundary contributions that contaminated the long-time tail of the reconstructed relaxation modulus. \emph{i\text{-}Rheo-Tempo} avoids this problem by treating the zero-frequency limit explicitly. In its most general form, the value of $\eta'(0)$ is estimated from a local fit of the low-frequency experimental spectrum, while $\eta''(0)$ is imposed either as zero (the default condition for viscoelastic fluids) or as a user-defined value. The point
\begin{equation}
\bigl(\omega,\eta'(\omega),\eta''(\omega)\bigr)
=
\bigl(0,\eta'(0),\eta''(0)\bigr)
\end{equation}
is then inserted as the first node of the spectrum before interpolation. In this way, the low-frequency boundary enters the algorithm as an explicit physical \textit{anchor} rather than as an artificial extrapolated branch.

A more physically constrained option can also be adopted in the terminal regime. Over a user-selected low-frequency range, the experimental dynamic moduli may be fitted with a single Maxwell mode,
\begin{align}
G'(\omega) &= \frac{G_0(\omega\tau)^2}{1+(\omega\tau)^2},
\label{eq:maxwell_gp}
\\
G''(\omega) &= \frac{G_0(\omega\tau)}{1+(\omega\tau)^2},
\label{eq:maxwell_gpp}
\end{align}
where $G_0$ and $\tau$ denote the modulus and relaxation time of the terminal mode, respectively. The corresponding viscosity components are
\begin{align}
\eta'(\omega)
&=
\frac{G''(\omega)}{\omega}
=
\frac{G_0\tau}{1+(\omega\tau)^2},
\label{eq:maxwell_eta1}
\\
\eta''(\omega)
&=
\frac{G'(\omega)}{\omega}
=
\frac{G_0\tau^2\,\omega}{1+(\omega\tau)^2}.
\label{eq:maxwell_eta2}
\end{align}
Hence,
\begin{equation}
\eta'(0)=G_0\tau,
\qquad
\eta''(0)=0.
\label{eq:maxwell_zero_limit}
\end{equation}
When this option is selected, the fitted Maxwell mode provides the low-frequency extension down to $\omega=0$, thereby enforcing the correct terminal scaling $G'(\omega)\sim\omega^2$ and $G''(\omega)\sim\omega$ and, consequently, a physically consistent long-time tail of $G(t)$.
Notice that, the optional Maxwell-tail extension is introduced purely as a diagnostic and regularising aid for the terminal regime; it is not part of the core inversion algorithm and does not influence the reconstructed $G(t)$ within the experimentally supported time window, affecting only the asymptotic long-time behaviour beyond the reliable data range.

At high frequencies, by contrast, a modest local completion remains useful. The experimental $\eta'(\omega)$ and $\eta''(\omega)$ are fitted near $\omega_{\max}^{\mathrm{exp}}$ using low-order polynomials in $\log\omega$, and the spectrum is extended over a limited additional range. This extension is not intended to assign physical meaning to the unmeasured region, but only to regularise the interpolation and suppress spurious curvature at the upper boundary.

The interpolation itself is performed in linear $\omega$ space on the dataset composed of the explicit zero-frequency point, the experimental spectrum, and the local high-frequency extension. The resulting interpolant is then resampled on a logarithmically spaced frequency grid,
thereby providing a numerically well-conditioned representation across the full experimentally relevant bandwidth while preserving the natural variable with respect to which the derivatives are taken.

Importantly, the extended and resampled spectrum is used only to stabilise the inversion; no physical interpretation is ascribed to the completed region outside the measured domain. Accordingly, the
reconstructed relaxation modulus is reported exclusively within the reciprocal experimental window
\begin{equation}
t\in
\left[
\frac{1}{\omega_{\max}^{\mathrm{exp}}},
\frac{1}{\omega_{\min}^{\mathrm{exp}}}
\right],
\end{equation}
so that all quoted values of $G(t)$ remain directly supported by experimental data. The boundary-conditioning step therefore removes numerical artefacts without constituting an extrapolation of the material response beyond the measured rheological window.

\subsection{Final Reconstruction, Error Scaling, and Interpolation Sensitivity}

In the practical implementation of \emph{i\text{-}Rheo-Tempo}, the interval formulation is not applied directly to the discrete experimental spectrum. Instead, the experimental complex-viscosity data are first used to construct a continuous representation of the spectrum. More specifically, the experimental values
$\{\omega_k,\eta'(\omega_k),\eta''(\omega_k)\}$ are augmented with an explicit zero-frequency point and a modest high-frequency extension.
The resulting dataset is then interpolated in $\omega$ using shape-preserving methods and subsequently resampled on a logarithmically spaced frequency grid spanning the interval $\omega \in [0,\omega_{\max}^{\mathrm{ext}}]$.

The slopes $a_k$ entering the interval formulation are therefore computed from this resampled spectrum rather than from the original experimental nodes. Consequently, the inversion effectively evaluates the inverse transform of the interpolated viscosity function defined on the extended domain. Because the resampled spectrum explicitly spans the finite interval $[0,\omega_{\max}^{\mathrm{ext}}]$, the boundary
integrals that appear in the formal derivation are naturally absorbed into the first and last frequency intervals of the  sum.
The inverse transform therefore reduces to a pure interval-slope formulation evaluated on the extended spectrum, namely
\begin{align}
G_{\eta'}(t)
&=
\frac{2}{\pi t^2}
\sum_{k=1}^{N-1}
a_k^{(1)}
\bigl[
\cos(\omega_{k+1} t)-\cos(\omega_{k} t)
\bigr],
\label{eq:G_eta1_final}
\\
G_{\eta''}(t)
&=
\frac{2}{\pi}
\left[
\frac{\eta''(0)}{t}
+
\frac{1}{t^2}
\sum_{k=1}^{N-1}
a_k^{(2)}
\bigl[
\sin(\omega_{k+1} t)-\sin(\omega_k t)
\bigr]
\right],
\label{eq:G_eta2_final}
\end{align}
where $\{\omega_k\}_{k=1}^{N}$ denotes the discrete set of angular frequencies defining the interpolated spectrum.
For viscoelastic fluids, where $\eta''(0)=0$, both branches are therefore evaluated entirely from the interval slopes of the interpolated viscosity spectrum.

\paragraph*{Error scaling.} At long times, the reconstruction of $G(t)$ becomes controlled almost entirely by the lowest-frequency intervals of the spectrum, where the dominant contributions arise from the first one or two low-frequency bins, for which $\omega_k t \ll 1$. Expanding
the kernels for small arguments gives
\begin{equation}
\frac{1}{t^2}
\left[
\cos(\omega_k t)-\cos(\omega_{k+1} t)
\right]
\approx
\frac{1}{2}\left(\omega_{k+1}^2-\omega_k^2\right),
\label{eq:cosine_small_argument}
\end{equation}
and
\begin{equation}
\frac{1}{t^2}
\left[
\sin(\omega_{k+1} t)-\sin(\omega_k t)
\right]
\approx
\frac{\omega_{k+1}-\omega_k}{t}.
\label{eq:sine_small_argument}
\end{equation}
Accordingly, the leading asymptotic contribution of the $k$th interval
scales as
\begin{equation}
\Delta G_{\eta'}^{(k)}(t)
\sim
a_k^{(1)}\,\omega_k\,\Delta\omega_k,
\qquad
\Delta G_{\eta''}^{(k)}(t)
\sim
a_k^{(2)}\,\frac{\Delta\omega_k}{t},
\label{eq:interval_scalings_longtime}
\end{equation}
with $\Delta\omega_k=\omega_{k+1}-\omega_k$. These relations make clear that the long-time tail is determined not by the absolute quality of the measured moduli in a visual sense, but by the accuracy with which the lowest-frequency slopes of the viscosity spectrum are estimated.
\begin{figure*}[!t]
    \centering
    \includegraphics[width=1\linewidth]{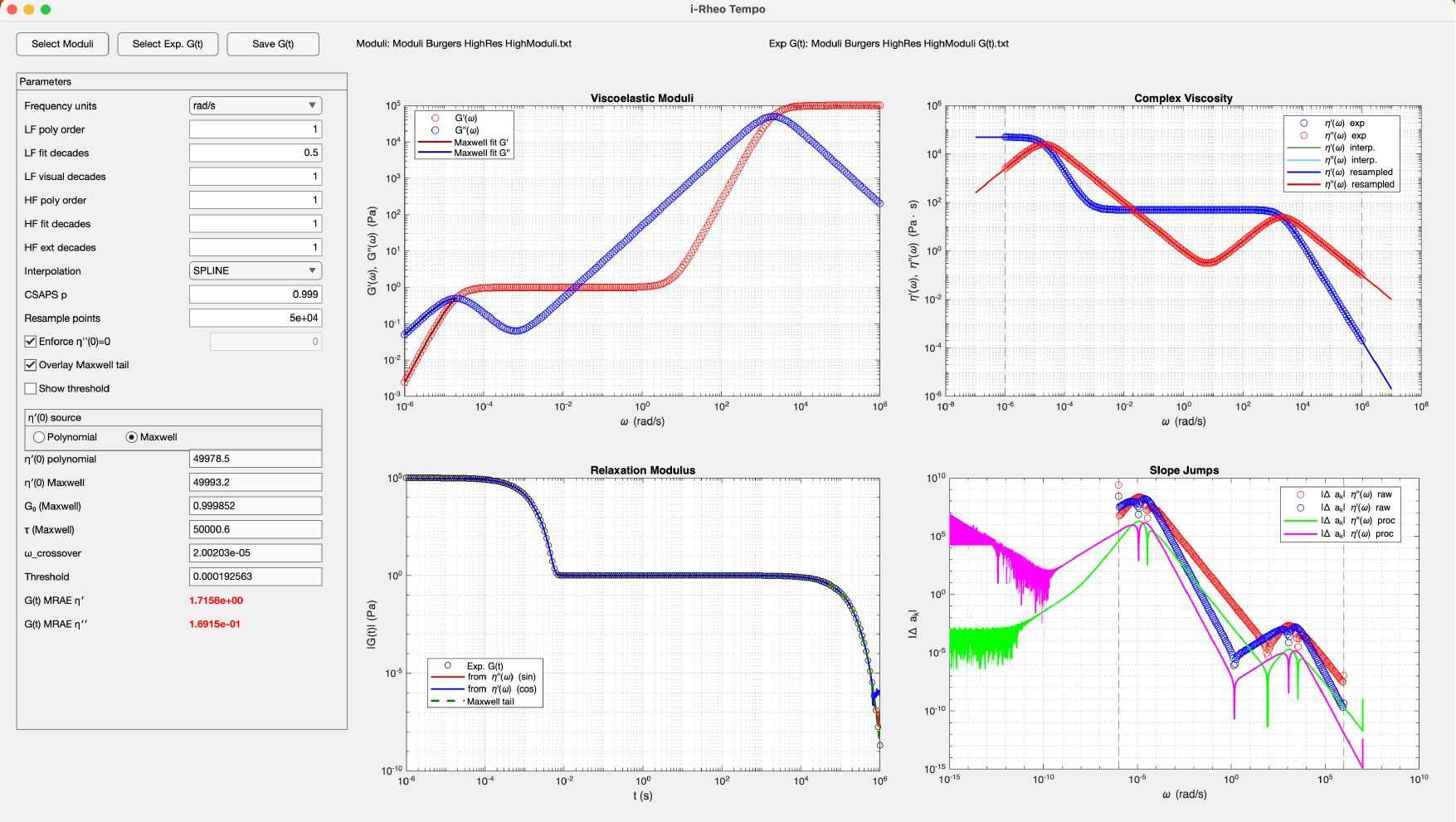}
\caption{\textbf{Graphical user interface (GUI) of \emph{i\text{-}Rheo-Tempo} (MATLAB implementation) applied to synthetic Burgers-model data.} The left panel allows loading of frequency-domain data and optional $G(t)$, selection of reconstruction parameters (low/high-frequency conditioning, interpolation, resampling, and boundary constraints such as $\eta'(0)$ enforcement and Maxwell-tail overlay), and displays key outputs. The right panel shows diagnostic plots: \textbf{(top left)} $G'(\omega)$ and $G''(\omega)$ with optional Maxwell fit; \textbf{(top right)} $\eta'(\omega)$ and $\eta''(\omega)$ (experimental, interpolated, resampled); \textbf{(bottom left)} reconstructed $G(t)$ from both branches with optional reference data; \textbf{(bottom right)} slope-jump spectra for raw and processed data. MATLAB and Python implementations are available under DOI: \href{http://dx.doi.org/10.5525/gla.researchdata.2230}{10.5525/gla.researchdata.2230}.}
    \label{fig:GUI}
\end{figure*}
A useful practical consequence follows immediately. Once $t$ becomes comparable to $1/\omega_{\min}^{\mathrm{exp}}$, the reconstruction is dominated by the first low-frequency interval, and the tail becomes numerically fragile. Denoting by $\delta a_1$ the uncertainty in the slope of that interval, the corresponding asymptotic error floor scales as
\begin{equation}
|G(t)|
\lesssim
\frac{2}{\pi}\,
|\delta a_1|\,
\frac{\Delta\omega_1}{t},
\label{eq:error_floor_longtime}
\end{equation}
where $\Delta\omega_1=\omega_2-\omega_1$ denotes the width of the first experimental frequency interval, and this expression is the most restrictive of the two in Eq. \eqref{eq:interval_scalings_longtime}.

For practical diagnostics, this asymptotic estimate can be converted into a conservative constant threshold by evaluating the long-time criterion at the largest experimentally supported time, $t_{\max}=1/\omega_{\min}^{\mathrm{exp}}$. This yields
\begin{equation}
G_{\mathrm{thr}}
\sim
\frac{2}{\pi}\,
|\delta a_1|\,
\Delta\omega_1\,
\omega_{\min}^{\mathrm{exp}},
\label{eq:constant_threshold}
\end{equation}
which provides a simple reference scale for identifying the onset of numerical fragility in the reconstructed relaxation modulus.

\paragraph*{Dependence on interpolation and resampling.}
Although the present formulation is model-free in the sense that it does not impose any predefined relaxation spectrum, its numerical implementation involves interpolation and resampling steps that may, in principle, influence the local curvature of the spectrum and therefore the reconstructed $G(t)$. 
In practice, the inversion is robust with respect to these algorithmic choices when the experimental dataset is sufficiently dense. In this regime, interpolation schemes that preserve smoothness and avoid spurious oscillations, combined with logarithmic resampling on adequately dense grids, yield consistent reconstructions across all datasets considered. 

However, when the density of experimental data points is low, the choice of interpolation scheme becomes critical. In such cases, different interpolation methods can introduce noticeable differences in the reconstructed $G(t)$, reflecting their distinct constraints on local curvature and higher-order derivatives. This behaviour is consistent with previous findings on Fourier-transform-based rheological analysis, where the fidelity of interpolation strongly depends on the density of initial sampling points and the presence of noise~\cite{smith2021}. In particular, shape-preserving schemes (e.g.\ PCHIP) tend to provide more stable reconstructions at low data density by avoiding overshoots, whereas spline-based methods may introduce artificial oscillations.

Systematic tests performed by varying interpolation methods and resampling densities show that these differences remain confined to regimes where the information content of the data is intrinsically limited. Within the experimentally supported time window, all methods converge when sufficient data density is available, while discrepancies become apparent only at long times or under sparse sampling conditions. These observations further support the interpretation that long-time fluctuations primarily reflect finite spectral bandwidth and limited data resolution, rather than the specific numerical implementation.


\section{Test Cases}

\subsection{Analytical Benchmarking with Multi-Mode Viscoelastic Models}

Prior to analysing experimental datasets, the performance of \emph{i\text{-}Rheo-Tempo} is first assessed using synthetic dynamic-moduli data generated from a two-mode Burgers model, as reported in Figure~\ref{fig:GUI}. The model parameters are deliberately selected to mimic the complex viscoelastic response of polymeric melts, characterised by a pronounced separation between elastic and viscous contributions and a broad distribution of relaxation times. As a result, the frequency spectrum exhibits two main relaxation components well separated in time, and non-trivial slope variations, thereby providing a stringent benchmark for the interval-slope inversion. Owing to the availability of the exact analytical form of the relaxation modulus, this synthetic dataset offers a fully controlled framework to evaluate the accuracy, robustness, and numerical consistency of the reconstruction, independently of experimental noise or measurement artefacts.
\begin{figure}[!t]
    \centering
    \includegraphics[width=1\linewidth]{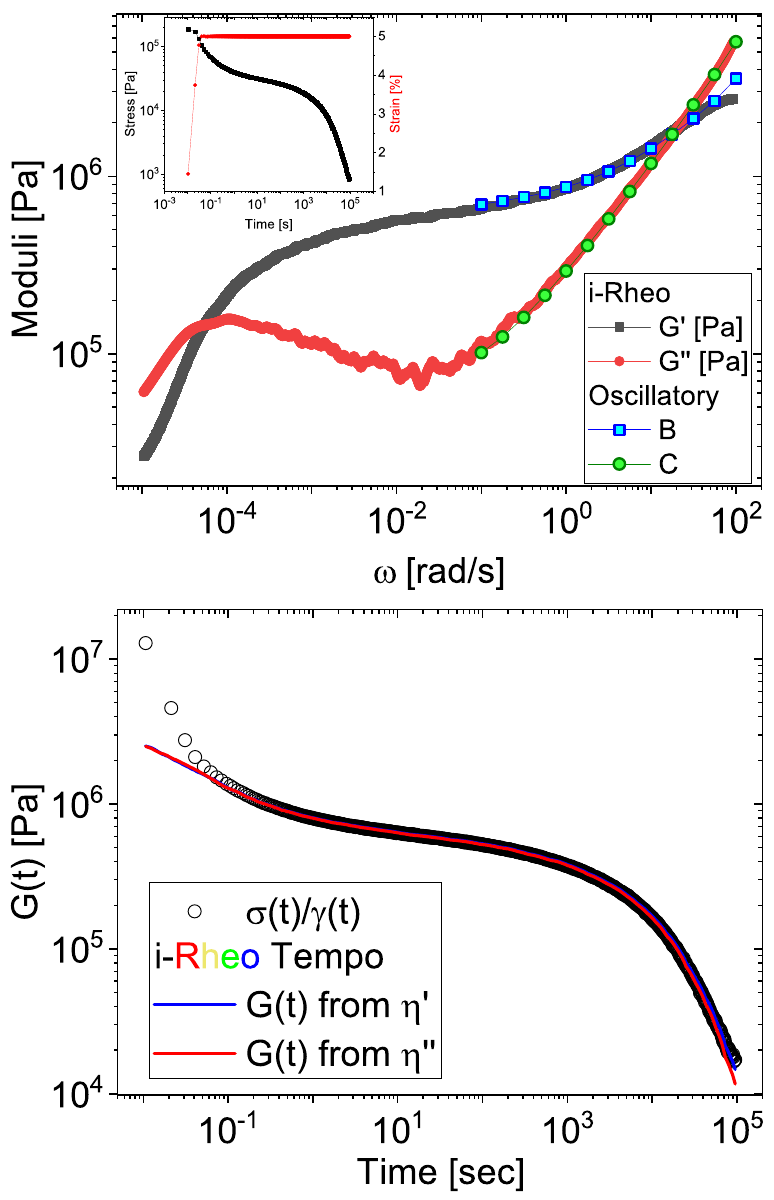}
    \caption{\textbf{Application of \emph{i\text{-}Rheo-Tempo} to an industrial styrene–butadiene rubber (SBR).} \textbf{(Top)} Frequency-domain viscoelastic moduli of SBR: storage $G'$ and loss $G''$ obtained via \emph{i-Rheo} from step-strain data (inset), compared with oscillatory measurements\cite{tassieri2016}. \textbf{(Bottom)} Time-domain relaxation modulus $G(t)$ reconstructed using \emph{i\text{-}Rheo-Tempo} against the direct estimate $\sigma(t)/\gamma(t)$.}
    \label{fig:iSBR}
\end{figure}
An important validation of the method emerges from the reversibility of the transform, illustrated using a two-mode Burgers model similar to the previous one, but with an extended terminal regime, as reported in Figure~\ref{fig:ClosedLoop} in the Appendix~\ref{appendixA}. Although the reconstructed relaxation modulus $G(t)$ may exhibit apparent oscillations or noise at very long times---arising from the amplification of low-frequency uncertainties inherent to the second-derivative formulation---these features do not carry additional physical information. Indeed, when the reconstructed $G(t)$ is transformed back into the frequency domain using \emph{i\text{-}Rheo-\textit{GT}}~\cite{tassieri2018}, a tool that is based on the same analytical framework, the original dynamic moduli are recovered within the experimental window without distortion. This demonstrates that the forward and inverse transforms are internally consistent and numerically stable, and that the long-time fluctuations originate from the finite spectral bandwidth rather than from a loss of physical fidelity.
\begin{figure}[t]
    \centering
    \includegraphics[width=1\linewidth]{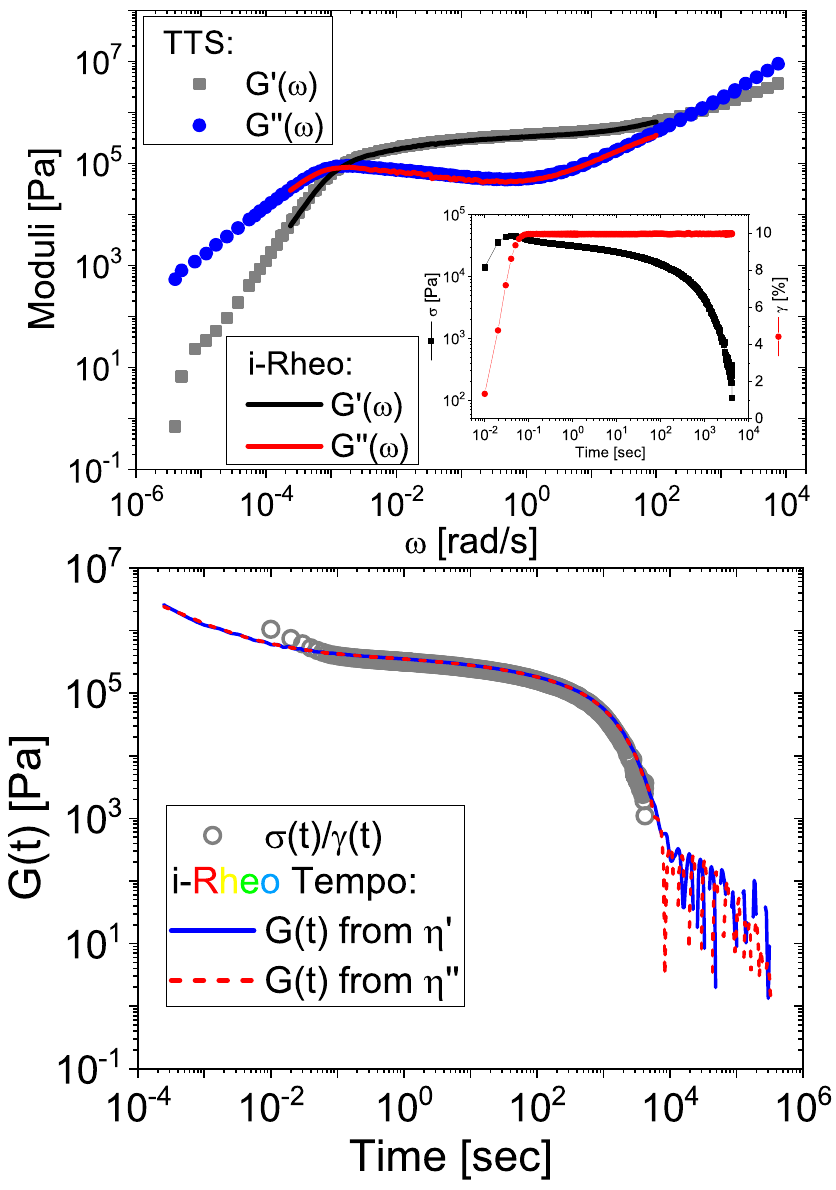}
    \caption{\textbf{Application of \emph{i\text{-}Rheo-Tempo} to a monodisperse linear polyisoprene melt.} \textbf{(Top)} Dynamic moduli $G'(\omega)$ and $G''(\omega)$ obtained via time--temperature superposition (symbols)~\cite{auhl2008linear}, compared with those reconstructed using \emph{i\text{-}Rheo}~\cite{tassieri2016} from raw step-strain data (inset). \textbf{(Bottom)} Relaxation modulus $G(t)$ recovered from the TTS-based frequency-domain data using \emph{i\text{-}Rheo-Tempo}, showing excellent agreement with the experimental estimate $\sigma(t)/\gamma(t)$ over an extended time window. The apparent fluctuations at long times—emerging approximately one decade beyond the inverse of the low-frequency crossover—originate from the amplification of low-frequency uncertainties and do not carry additional physical information.}
    \label{fig:PI}
\end{figure}
\subsection{Industrial styrene–butadiene rubber}
In order to corroborate the performance of \emph{i\text{-}Rheo-Tempo}, we revisited one of the benchmark datasets originally analysed in 2016 using \emph{i\text{-}Rheo}, namely the stress–relaxation measurement of an industrial styrene–butadiene rubber (SBR) \cite{tassieri2016}. SBR is a widely used elastomer whose relaxation dynamics reflect a broad distribution of relaxation modes arising from its mixed linear/branched microstructure and the chemical heterogeneity of styrene–butadiene random copolymers. When the output of \emph{i\text{-}Rheo} applied to the experimental step-strain dataset is processed with \emph{i\text{-}Rheo-Tempo}, the reconstructed relaxation modulus \(G(t)\) shows a good agreement with the experimental values obtained directly as the ratio between the measured stress and strain, as reported in Figure~\ref{fig:iSBR}. The only visible deviation––a small mismatch at very short times––is not physically concerning: the constitutive relation linking stress, strain and the relaxation modulus is given by the convolution integral:
\begin{equation}
\sigma(t) = \int_{0}^{t} G(t - t')\,\dot{\gamma}(t')\,\mathrm{d}t',
\label{eq:conv_integ}
\end{equation}
and therefore the instantaneous estimate \(G(t) \approx \sigma(t)/\gamma(t)\) is not exact, especially at early times when \(\dot{\gamma}(t)\) is large and the deformation ramp is still evolving. Despite this subtlety, the reconstructed \(G(t)\) overlaps the experimental relaxation curve over the entire experimentally supported time window, demonstrating the accuracy, robustness, and practical reliability of the new inversion framework.

\subsection{Monodisperse linear polyisoprene melt}

As a further validation of \emph{i\text{-}Rheo-Tempo}, we revisit the case of a monodisperse linear polyisoprene melt, a canonical system extensively studied in polymer rheology. Owing to their simple topological structure---namely, entangled linear chains---such systems have played a central role in the development and validation of constitutive theories based on the tube model framework. From an experimental standpoint, their linear viscoelastic (LVE) properties are typically accessed over broad frequency ranges via time--temperature superposition (TTS), which combines measurements performed at different temperatures into a single master curve.
\begin{figure}[t]
    \centering
    \includegraphics[width=1\linewidth]{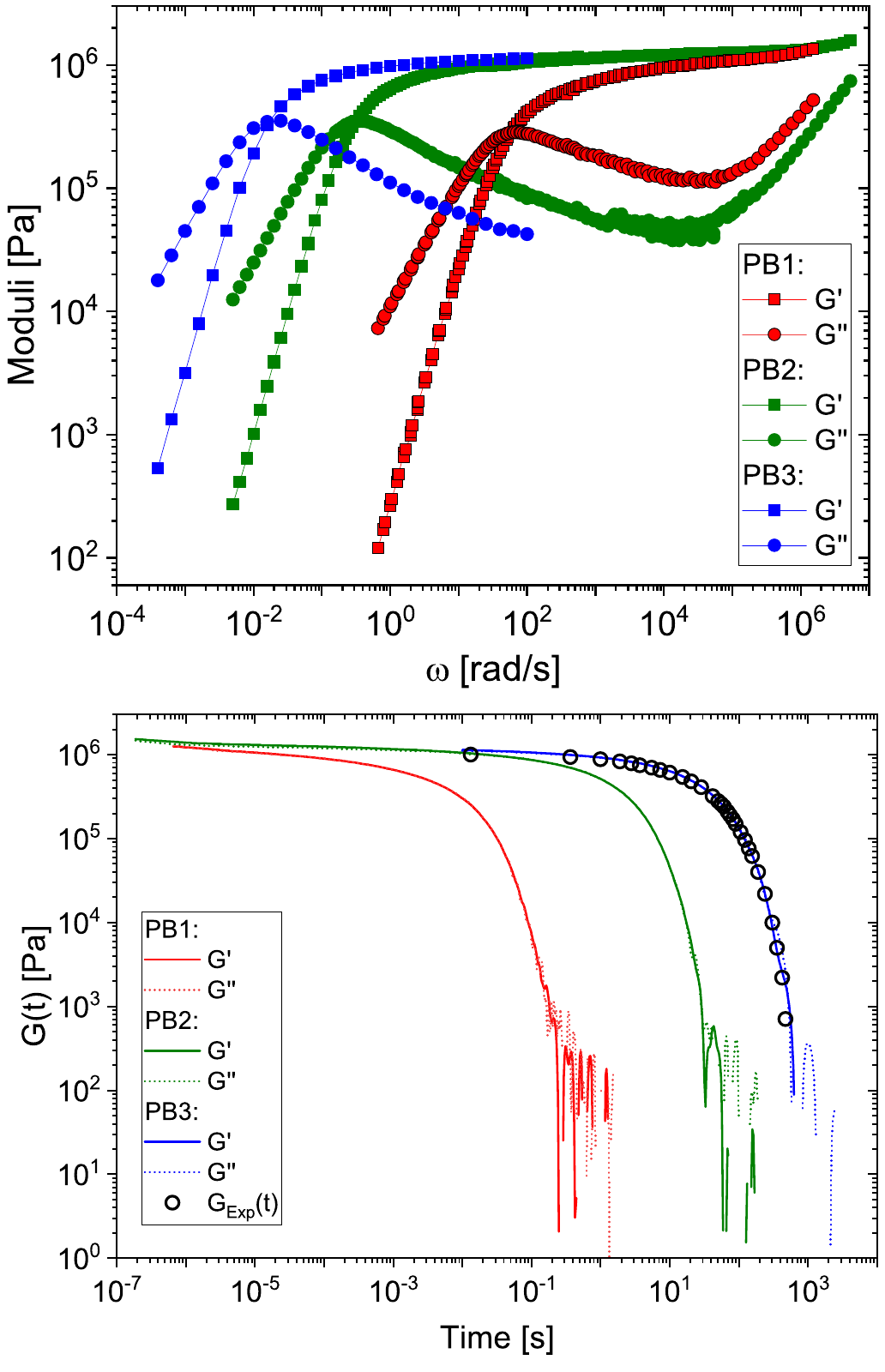}
 \caption{\textbf{Application of \emph{i\text{-}Rheo-Tempo} to monodisperse linear polybutadiene melts.} \textbf{(Top)} Dynamic moduli $G'(\omega)$ and $G''(\omega)$ for three samples with different molecular weights (PB1--PB3). \textbf{(Bottom)} Relaxation modulus $G(t)$ reconstructed using \emph{i\text{-}Rheo-Tempo} for all samples, compared with the experimental estimate $G_{\mathrm{exp}}(t)=\sigma(t)/\gamma(t)$ available for PB3, showing excellent agreement over the accessible time window. At short times, the reconstructed $G(t)$ for all samples collapses onto a common curve, as expected since the dynamics at these time scales are governed by local inter- and intra-molecular interactions, which are independent of molecular weight.}
    \label{fig:RC}
\end{figure}
Here, rather than relying on step-strain data, we directly exploit TTS-based oscillatory measurements as input to \emph{i\text{-}Rheo-Tempo}, thereby reconstructing the relaxation modulus $G(t)$ over an extended time window. This represents a complementary use of the method, demonstrating its applicability beyond single-measurement protocols and its ability to extract time-domain information from broadband frequency-domain datasets.

As shown in Fig.~\ref{fig:PI}, the reconstructed $G(t)$ exhibits quantitative agreement with the experimental relaxation modulus obtained from the ratio $\sigma(t)/\gamma(t)$ over the entire time range supported by the data. This confirms both the accuracy and robustness of the inversion procedure when applied to high-quality TTS datasets.

At very long times, the reconstructed $G(t)$ displays apparent oscillations or noise, which typically emerge approximately one decade beyond the inverse of the low-frequency crossover. As discussed earlier, these features originate from the amplification of low-frequency uncertainties inherent to the second-derivative formulation and do not carry additional physical meaning. Importantly, they do not affect the fidelity of the reconstruction within the experimentally accessible window, nor do they compromise the physical interpretation of the material response.

\subsection{Monodisperse linear polybutadiene melts}

We next examine well-entangled linear polymer melt data inspired by Rubinstein and Colby~\cite{rubinstein1988,rubinstein2003}, consisting of three highly monodisperse polybutadiene samples with different molecular weights. These datasets (PB1, PB2, and PB3) correspond to samples with weight-average molecular weights of $M_w = 7.09 \times 10^4$, $3.55 \times 10^5$, and $9.25 \times 10^5$ g\,mol$^{-1}$, respectively, all characterised by a very narrow molecular weight distribution ($M_w/M_n < 1.1$) and identical microstructure within experimental uncertainty~\cite{rubinstein1988}.

While PB1 and PB2 exhibit extended oscillatory measurements with a well-defined rubbery plateau in $G'$ (top panel of Fig.~\ref{fig:RC}), all three systems display the hallmark features of entangled polymer dynamics, including a clear transition to the terminal regime consistent with reptation theory for highly monodisperse melts. These characteristics make them a stringent benchmark for assessing the performance of the inversion procedure over a wide range of time scales.

The dynamic moduli are used as input for \emph{i\text{-}Rheo-Tempo} to reconstruct the relaxation modulus $G(t)$ over an extended time window (bottom panel of Fig.~\ref{fig:RC}). As shown, the reconstructed $G(t)$ for all three samples captures the expected viscoelastic behaviour across the full time range, and is in good agreement with the experimental relaxation data $G_{\mathrm{exp}}(t)$ available for PB3, confirming the accuracy and robustness of the inversion procedure across different molecular weights. Notably, at short times the reconstructed $G(t)$ for all samples collapses onto a common curve, as expected since the dynamics at these time scales are governed by local inter- and intra-molecular interactions, which are independent of molecular weight.
\begin{figure}[!t]
    \centering
    \includegraphics[width=1\linewidth]{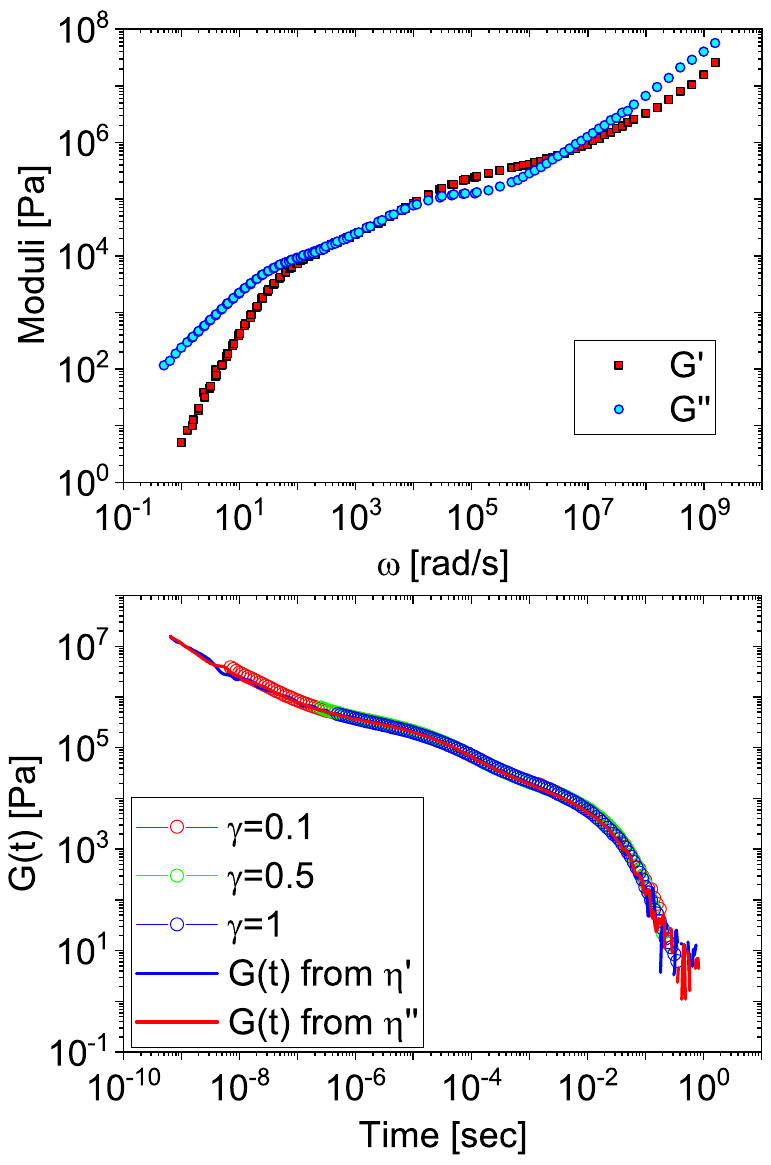}
\caption{\textbf{Application of \emph{i\text{-}Rheo-Tempo} to a model comb polymer (lc3-PBd1k at $\phi=0.5$).} \textbf{(Top)} Dynamic moduli obtained via time--temperature superposition~\cite{kapnistos2009}. \textbf{(Bottom)} Relaxation modulus $G(t)$ reconstructed using \emph{i\text{-}Rheo-Tempo}, compared with experimental stress--relaxation data measured at three strain amplitudes ($\gamma=0.1,\,0.5,$ and $1$)~\cite{kapnistos2009}, all within the linear viscoelastic regime. The reconstruction shows excellent agreement over the entire accessible time window, accurately capturing both the early-time branch relaxation and the subsequent slower backbone dynamics characteristic of comb architectures.}
    \label{fig:lc3c50}
\end{figure}
\subsection{Model comb polymers melt}
We further analyse a model comb polymer system reported by Kapnistos and Vlassopoulos~\cite{kapnistos2009}, focusing here on the sample lc3-PBd1k at $\phi=0.5$ (polybutadiene combs with $M_{backbone}$ = 50 kg/mol and 17 branches of 7 kg/mol, at volume fraction 0.5). This nearly monodisperse branched architecture exhibits hierarchical relaxation arising from distinct molecular mechanisms, namely branch retraction at short times and backbone reptation at longer times. As a result, the stress relaxation modulus displays a characteristic multi-step decay spanning several decades in time, making this system an ideal benchmark for testing the robustness of the inversion procedure across multiple relaxation modes.

The TTS-derived dynamic moduli (top panel of Fig.~\ref{fig:lc3c50}) are used as input for \emph{i\text{-}Rheo-Tempo} to reconstruct the relaxation modulus $G(t)$ (bottom panel). The reconstructed curves are compared with the experimentally measured $G(t)$ obtained from step-strain experiments at three different strain amplitudes ($\gamma = 0.1,\,0.5,$ and $1$)~\cite{kapnistos2009}, all within the linear viscoelastic regime. As shown, the \emph{i\text{-}Rheo-Tempo} reconstruction is in very good agreement with the experimental data over the entire accessible time window (spanning ten decades), accurately capturing both the early-time branch relaxation and the subsequent slower backbone dynamics. The consistency of the reconstructed $G(t)$ across nearly nine decades further confirms the reliability of the method and its ability to faithfully recover the underlying linear viscoelastic response from complex frequency-domain data, even in systems exhibiting strongly hierarchical relaxation mechanisms.

\begin{figure}[t]
    \centering
    \includegraphics[width=1\linewidth]{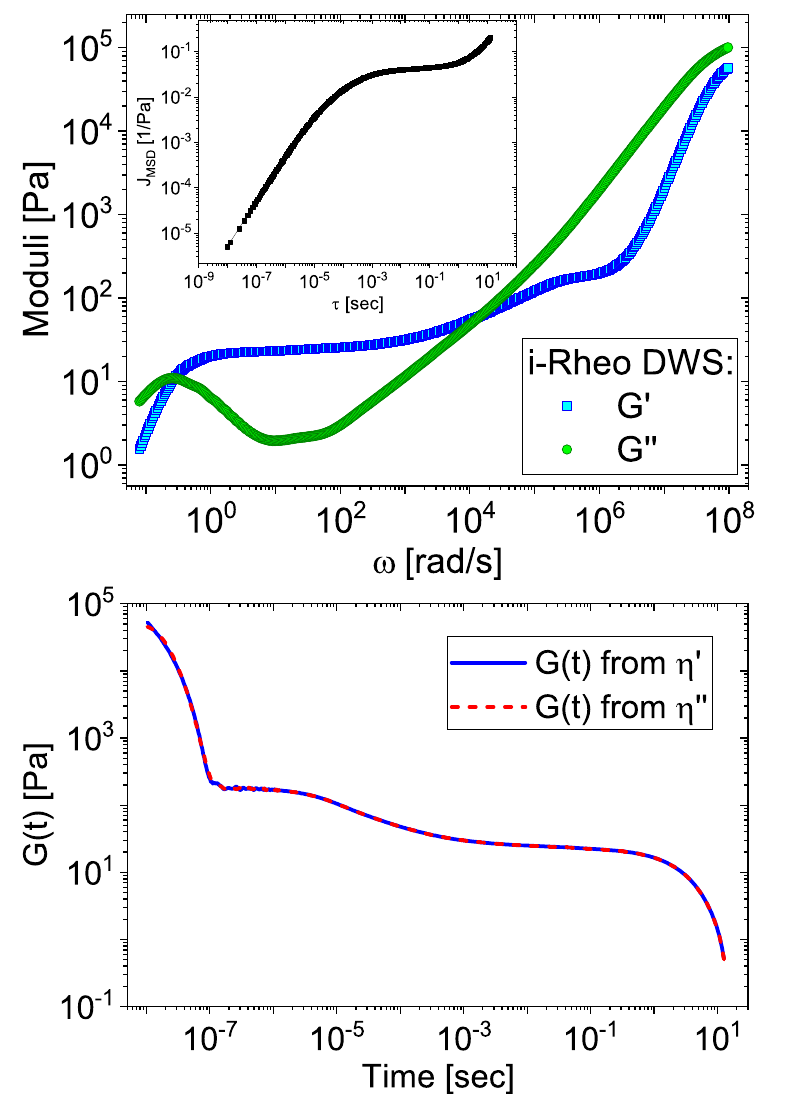}
\caption{\textbf{Application of \emph{i\text{-}Rheo-Tempo} to broadband DWS microrheology data.} \textbf{(Top)} Dynamic moduli $G'(\omega)$ and $G''(\omega)$ obtained from DWS measurements (symbols), spanning nearly nine decades in frequency. Inset: corresponding creep compliance derived from the mean-squared displacement. \textbf{(Bottom)} Relaxation modulus $G(t)$ reconstructed using \emph{i\text{-}Rheo-Tempo}, showing excellent agreement between the two independent reconstructions obtained from $\eta'(\omega)$ and $\eta''(\omega)$. The resulting $G(t)$ is smooth and physically consistent over the entire accessible time window, demonstrating the robustness of the method under broadband, inertia-affected conditions.}
    \label{fig:DWS}
\end{figure}

\subsection{Broadband microrheology: diffusing-wave spectroscopy}

Finally, we apply the method to broadband diffusing-wave spectroscopy (DWS) microrheology data reported by Scheffold and co-workers~\cite{Scheffold2026}. These measurements span nearly nine decades in frequency, probing regimes where inertia, experimental noise, and boundary artefacts can significantly affect the reliability of the extracted viscoelastic response.

The dynamic moduli shown in the top panel of Fig.~\ref{fig:DWS} were obtained using an adapted version of the model-free analytical method introduced by Tassieri \emph{et al.}~\cite{tassieri2012} for microrheology with optical tweezers, and subsequently used by Scheffold \emph{et al.}~\cite{Scheffold2026} to benchmark their model-based analysis. Here, they are used as input for \emph{i\text{-}Rheo-Tempo} to reconstruct the relaxation modulus $G(t)$ (bottom panel). As shown, the two independent reconstructions—obtained from $\eta'(\omega)$ and $\eta''(\omega)$—are in good agreement over the entire accessible time window, yielding a smooth and physically consistent $G(t)$ without the need for any fitting or regularisation.

Importantly, the reconstructed $G(t)$ captures the full hierarchy of dynamical regimes encoded in the broadband microrheology data, from the short-time, high-frequency behaviour—where inertia and hydrodynamic effects become relevant—to the long-time viscoelastic response of the material. The absence of spurious features and the near-perfect overlap between the two reconstructions provide a stringent validation of the method, demonstrating its numerical stability and robustness even when applied to state-of-the-art datasets spanning nearly ten decades in frequency.

This example highlights the capability of \emph{i\text{-}Rheo-Tempo} to reliably bridge frequency- and time-domain representations under the most demanding experimental conditions, effectively extending its applicability to modern broadband microrheology and reinforcing its role as a general framework for viscoelastic data analysis.

\subsection{Beyond rheology: a general framework for complex response functions}

Although the methodology has been introduced and validated here in the context of viscoelasticity, its applicability is not restricted to rheological systems. The \emph{i\text{-}Rheo-Tempo} framework operates on a complex-valued function of frequency, $F(\omega) = F'(\omega) + iF''(\omega)$, and reconstructs its corresponding time-domain representation through a direct analytical inversion. As such, it is fundamentally a general tool for the analysis of linear response functions.

The only requirements for the applicability of the method are that the real and imaginary components of $F(\omega)$ are sufficiently smooth (piecewise differentiable), consistent with causality, and defined over a finite frequency window with well-behaved boundary conditions. Under these conditions, the inversion yields a time-domain kernel that is physically meaningful within the experimentally accessible range, with any deviations at the boundaries arising solely from finite bandwidth effects.

This perspective places \emph{i\text{-}Rheo-Tempo} within a broader class of analytical tools aimed at bridging frequency- and time-domain representations of complex systems. In this sense, the method provides a model-free route to recover the temporal response of a system directly from spectral measurements, without imposing any a priori functional form.

While viscoelasticity offers a natural and well-established testbed—owing to the direct relation between $G^*(\omega)$ and $G(t)$—the same framework can, in principle, be extended to other fields where complex response functions arise. Representative examples include the complex dielectric permittivity $\varepsilon^*(\omega)$ in dielectric spectroscopy~\cite{Debye}, the electrical impedance $Z^*(\omega)$ in electrochemical systems~\cite{macdonald1990review}, and the optical susceptibility $\chi^*(\omega)$ probed in optical Kerr effect spectroscopy~\cite{zhong2008optical}, all of which share the same underlying structure of a causal complex response function. This opens the possibility of applying the method to a wide range of systems, from soft matter and biological materials to electronic and photonic devices, wherever a causal complex response function can be experimentally determined.

\section{Conclusions}

\emph{i\text{-}Rheo-Tempo} provides a closed-form, model-free solution to the frequency-to-time inverse problem in linear viscoelasticity. By reformulating the inverse Fourier transform in terms of the second derivatives of the complex viscosity and exploiting their exact distributional discretisation, the method eliminates numerical quadrature and enables a direct analytical reconstruction of the relaxation modulus from experimental data. It represents the first derivative-based inversion from the frequency to time domains within the \emph{i\text{-}Rheo} framework, closing the loop between frequency and time in rheology. 

The resulting inversion is expressed in a compact interval-slope formulation, in which each frequency interval contributes through the difference of neighbouring harmonic kernels weighted by the local spectral slope. In this representation, boundary contributions are naturally incorporated within the first and last intervals of the spectrum, thereby avoiding artefacts associated with artificial endpoint extrapolations.

To address the intrinsic limitations imposed by finite experimental bandwidth, the method incorporates an explicit boundary-conditioning strategy. The zero-frequency limit is introduced as a physical anchor, either through local fitting or via a Maxwell representation of the terminal regime, while a modest high-frequency completion is used solely to stabilise the numerical implementation. The reconstructed relaxation modulus is then reported strictly within the reciprocal experimental window, ensuring that all results remain directly supported by measured data.

The robustness and accuracy of the framework have been demonstrated across a wide range of systems, including synthetic models, polymer melts, industrial elastomers, comb polymers with hierarchical relaxation, and broadband microrheology datasets spanning nearly ten decades in frequency. In all cases investigated here, the method consistently recovers the correct time-domain behaviour without the need for fitting or model assumptions.

Beyond its application to viscoelasticity, the underlying formulation is inherently general. \emph{i\text{-}Rheo-Tempo} operates on any complex-valued function of frequency whose real and imaginary components are sufficiently smooth and consistent with causality. As such, it provides a general analytical framework for reconstructing time-domain responses from frequency-domain measurements, with potential applications extending to dielectric spectroscopy, electrical impedance analysis, optical response measurements, and other fields governed by linear response theory.

In this broader context, \emph{i\text{-}Rheo-Tempo} establishes a direct and model-independent route for extracting temporal dynamics from spectral data, effectively bridging frequency and time domains under realistic experimental conditions.

\begin{acknowledgments}
In memory of Tom McLeish.

We are grateful to Mike Evans, Dietmar Auhl, Dan Curtis, Eky Febrianto, and Anna Rył for valuable discussions and insightful comments.
We sincerely thank Ralph Colby, Dimitris Vlassopoulos, and Frank Scheffold for kindly sharing their experimental data, which were essential for the validation of this work.
\end{acknowledgments}

\section*{Data Availability Statement}

The data that support the findings of this study are available from the University of Glasgow Enlighten: Research Data repository under the DOI \href{http://dx.doi.org/10.5525/gla.researchdata.2230}{10.5525/gla.researchdata.2230}. 

The repository includes all datasets used in this work, together with the MATLAB and Python implementations of the \emph{i\text{-}Rheo-Tempo} software required to reproduce the results.

\appendix
\begin{figure*}[!t]
    \centering
    \includegraphics[width=1\linewidth]{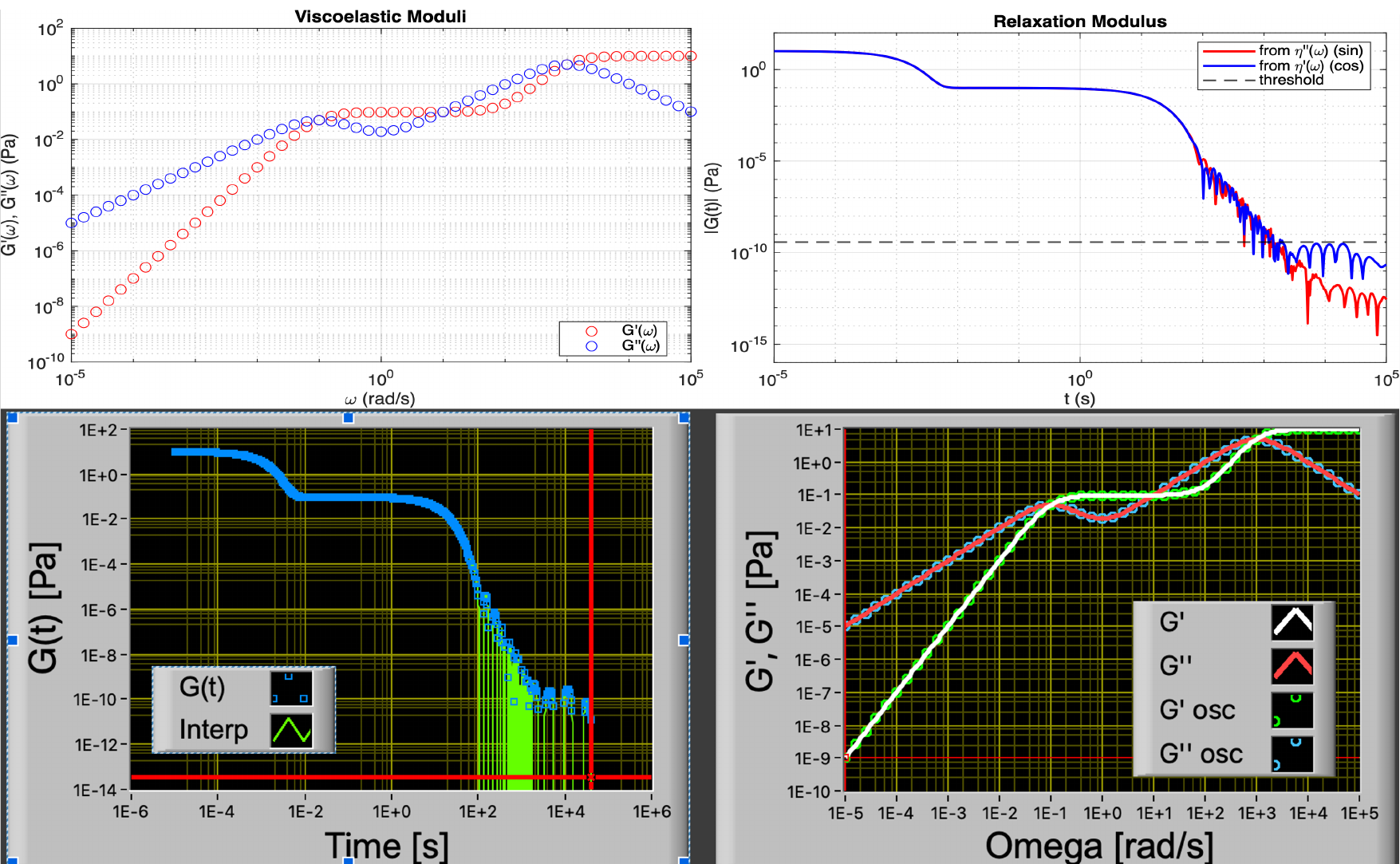}
    \caption{\textbf{Reversibility and internal consistency of the \emph{i\text{-}Rheo-Tempo} framework demonstrated on a two-mode Burgers model.} \textbf{(Top left)} Input dynamic moduli $G'(\omega)$ and $G''(\omega)$ from \emph{i\text{-}Rheo-Tempo}. \textbf{(Top right)} Relaxation modulus $G(t)$ reconstructed via \emph{i\text{-}Rheo-Tempo} from $\eta'(\omega)$ and $\eta''(\omega)$, showing apparent long-time oscillations due to low-frequency uncertainty amplification. \textbf{(Bottom left)} Same $G(t)$, but fed as input to \emph{i\text{-}Rheo-\textit{GT}}~\cite{tassieri2018}. \textbf{(Bottom right)} Dynamic moduli obtained by transforming the reconstructed $G(t)$ back to the frequency domain using \emph{i\text{-}Rheo-\textit{GT}}, recovering the original $G'(\omega)$ and $G''(\omega)$ within the experimental window. This confirms that the forward and inverse transforms are internally consistent and numerically stable, and that long-time fluctuations arise from finite spectral bandwidth rather than loss of physical fidelity.}
    \label{fig:ClosedLoop}
\end{figure*}
 \section{Closed-Loop Validation and Interpretation of Long-Time Behaviour}\label{appendixA}
 An important validation of the \emph{i\text{-}Rheo-Tempo} framework is provided by the reversibility of the transform, as illustrated in Fig.~\ref{fig:ClosedLoop} using a two-mode Burgers model with an extended terminal regime. Starting from the dynamic moduli (top left), the relaxation modulus $G(t)$ is reconstructed via \emph{i\text{-}Rheo-Tempo} (top right). As clearly visible, the reconstructed $G(t)$ remains smooth and physically consistent over the experimentally supported time window, while exhibiting apparent oscillations at long times, beyond the threshold indicated in the figure, in agreement with Equation~\eqref{eq:error_floor_longtime} in the main text.  
 These oscillations arise from the amplification of low-frequency uncertainties inherent to the second-derivative formulation and therefore reflect the finite spectral bandwidth rather than any physical feature of the material response.

To assess their physical significance, the reconstructed $G(t)$ is subsequently used as input for \emph{i\text{-}Rheo-\textit{GT}}~\cite{tassieri2018} (bottom left), performing the forward transformation back to the frequency domain. The resulting dynamic moduli (bottom right) recover the original $G'(\omega)$ and $G''(\omega)$ within the experimental window without distortion. This closed-loop consistency demonstrates that the forward and inverse transforms are internally consistent and numerically stable, and confirms that the long-time oscillations observed in $G(t)$ do not carry additional physical information.

Importantly, the same behaviour is consistently observed across all systems analysed in this work—including polymer melts, industrial elastomers, comb polymers, and broadband microrheology datasets—where analogous long-time fluctuations emerge beyond the reliable time window. In all cases, these features coincide with the onset of the threshold regime and therefore constitute a numerical signature of bandwidth limitation rather than a loss of physical fidelity.

\section{Graphical user interface (GUI) of \emph{i-Rheo-Tempo}}\label{appendixB}

\begin{figure*}[t]
    \centering
    \includegraphics[width=1\linewidth]{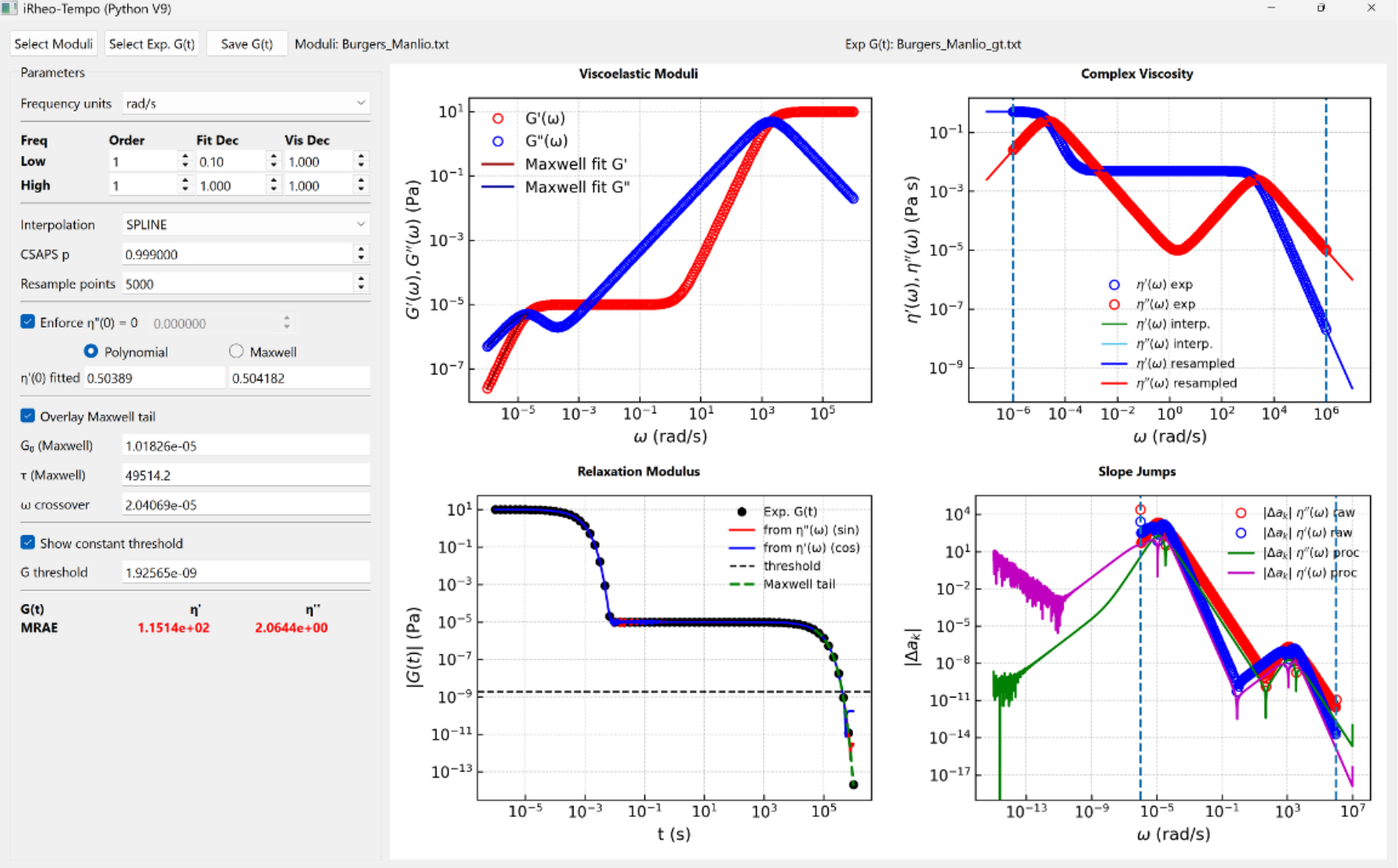}
\caption{\textbf{Graphical user interface (GUI) of \emph{i\text{-}Rheo-Tempo} (Python implementation) applied to synthetic Burgers-model data.} 
The left panel provides controls for loading frequency-domain data and optional $G(t)$, selecting reconstruction parameters (low- and high-frequency conditioning, interpolation method, resampling density, and boundary constraints such as enforcement of $\eta''(0)=0$ and optional Maxwell-tail overlay), and displays key numerical outputs and error metrics. 
The right panel presents diagnostic plots: \textbf{(top left)} storage and loss moduli $G'(\omega)$ and $G''(\omega)$ with optional Maxwell fit; \textbf{(top right)} complex viscosity components $\eta'(\omega)$ and $\eta''(\omega)$ (experimental, interpolated, and resampled); \textbf{(bottom left)} reconstructed relaxation modulus $G(t)$ from both branches compared with reference data and threshold; \textbf{(bottom right)} slope-jump spectra $|\Delta a_k|$ for raw and processed data, highlighting the curvature-based discretisation. 
MATLAB and Python implementations are available under DOI: \href{http://dx.doi.org/10.5525/gla.researchdata.2230}{10.5525/gla.researchdata.2230}.}
    \label{fig:GUIPython}
\end{figure*}

The interface comprises (left) a parameter panel with user inputs and read-only outputs, and (right) four diagnostic plots that document each stage of the reconstruction from frequency-domain data to the relaxation modulus \(G(t)\). Synthetic input data \(\{ \omega_k, G'(\omega_k), G''(\omega_k) \}\) (e.g. Burgers model) are loaded via \texttt{Select Moduli}; optionally, experimental \(G(t)\) can be overlaid via \texttt{Select Exp.\ \(G(t)\)}. The software converts the moduli into complex viscosity \(\eta^*(\omega)=\eta'(\omega)-i\,\eta''(\omega)=G^*(\omega)/(i\omega)\), applies boundary conditioning, constructs an interpolant in linear \(\omega\), resamples the spectrum on a logarithmic grid (including the explicit zero-frequency node), and evaluates \(G(t)\) through the interval-slope formulation of the second-derivative inversion.

\paragraph*{User input parameters (left panel).}
\begin{itemize}
  \item \emph{Frequency units}: \texttt{Hz} or \texttt{rad/s} (conversion \(\omega=2\pi f\) when required).
  \item \emph{Frequency range controls}: parameters defining the low- and high-frequency regions used for conditioning (e.g. number of decades and fitting order).
  \item \emph{Low-frequency conditioning}: estimation of \(\eta'(0)\) via local fitting of the low-frequency spectrum.
  \item \emph{High-frequency conditioning}: local fitting and modest extension of the spectrum near \(\omega_{\max}\) for numerical regularisation.
  \item \emph{Interpolation}: \texttt{PCHIP}, \texttt{SPLINE}, or \texttt{CSAPS} (smoothing spline).
  \item \emph{Smoothing parameter}: active only when smoothing splines are selected.
  \item \emph{Resampling density}: number of logarithmically spaced frequency points used in the inversion.
  \item \emph{Boundary constraint}: enforcement of \(\eta''(0)=0\) (default for viscoelastic fluids) or user-defined value.
  \item \emph{Optional Maxwell tail}: single-mode fit based on the lowest-frequency crossover \(G'=G''\), used as a terminal reference.
  \item \emph{Threshold display}: visualisation of the long-time robustness limit.
\end{itemize}

\paragraph*{Read-only outputs (left panel).}
\begin{itemize}
  \item \(\eta'(0)\) (fitted): value used to define the explicit \(\omega=0\) node.
  \item Maxwell parameters \((G_0,\tau)\): when the terminal fit is enabled.
  \item Crossover frequency: associated with the terminal regime.
  \item Threshold: estimate of the long-time numerical reliability limit.
  \item Error metrics (e.g. MRAE): quantitative indicators of reconstruction consistency between branches.
\end{itemize}

\paragraph*{Diagnostic plots (right panel).}
\begin{enumerate}
  \item[\textbf{(a)}] \textbf{Viscoelastic Moduli}—\(\log\!-\!\log\) plot of \(G'(\omega)\) and \(G''(\omega)\), with optional Maxwell fit.
  \item[\textbf{(b)}] \textbf{Complex Viscosity}—experimental, interpolated, and resampled \(\eta'(\omega)\) and \(\eta''(\omega)\); vertical markers indicate the experimental bandwidth \([\omega_{\min}^{\mathrm{exp}},\omega_{\max}^{\mathrm{exp}}]\).
  \item[\textbf{(c)}] \textbf{Relaxation Modulus}—reconstructed \(G(t)\) from both inversion branches, optionally compared with experimental data and augmented with a Maxwell reference and robustness threshold.
  \item[\textbf{(d)}] \textbf{Slope Jumps}—magnitude of the interval slope variations \(|\Delta a_k|\) for raw and processed spectra, illustrating the curvature-based discretisation underlying the method.
\end{enumerate}

\paragraph*{Workflow summary.}
The dynamic moduli are converted to complex viscosity, conditioned at low and high frequency, interpolated in linear \(\omega\), and resampled on a logarithmic grid. The relaxation modulus is then computed using the interval-slope formulation
\[
G(t)=\frac{2}{\pi}t^{-2}\sum_k a_k\,[K_k(t)-K_{k+1}(t)],
\]
with \(K=\cos\) for \(\eta'\) and \(K=\sin\) for \(\eta''\). The optional Maxwell fit provides a physically constrained representation of the terminal regime, while the threshold identifies the onset of the numerically fragile long-time regime.

\nocite{*}
\section*{bibliography}
\bibliography{aipsamp}

@PREAMBLE{
 "\providecommand{\noopsort}[1]{}" 
 # "\providecommand{\singleletter}[1]{#1}%" 
}

@article{schwarzl1975,
  title={Numerical calculation of stress relaxation modulus from dynamic data for linear viscoelastic materials},
  author={Schwarzl, FR},
  journal={Rheologica Acta},
  volume={14},
  number={7},
  pages={581--590},
  year={1975},
  publisher={Springer}
}

@article{evans2009,
  title={Direct conversion of rheological compliance measurements into storage and loss moduli},
  author={Evans, RML and Tassieri, Manlio and Auhl, Dietmar and Waigh, Thomas A},
  journal={Physical Review E—Statistical, Nonlinear, and Soft Matter Physics},
  volume={80},
  number={1},
  pages={012501},
  year={2009},
  publisher={APS}
}

@article{tassieri2016,
  title={i-Rheo: Measuring the materials' linear viscoelastic properties “in a step”!},
  author={Tassieri, Manlio and Laurati, Marco and Curtis, Dan J and Auhl, Dietmar W and Coppola, Salvatore and Scalfati, Andrea and Hawkins, Karl and Williams, Phylip Rhodri and Cooper, Jonathan M},
  journal={Journal of Rheology},
  volume={60},
  number={4},
  pages={649--660},
  year={2016},
  publisher={AIP Publishing}
}

@book{Ferry1980,
  author    = {Ferry, John D.},
  title     = {Viscoelastic Properties of Polymers},
  edition   = {3},
  year      = {1980},
  publisher = {John Wiley \& Sons},
  address   = {New York},
  isbn      = {978-0471048947}
}

@article{schwarzl1968,
  title={Analysis of relaxation measurements},
  author={Schwarzl, FR and Struik, LCE},
  journal={Advances in molecular relaxation processes},
  volume={1},
  number={3},
  pages={201--255},
  year={1968},
  publisher={Elsevier}
}

@article{tassieri2012,
  title={Microrheology with optical tweezers: data analysis},
  author={Tassieri, Manlio and Evans, RML and Warren, Rebecca L and Bailey, Nicholas J and Cooper, Jonathan M},
  journal={New Journal of Physics},
  volume={14},
  number={11},
  pages={115032},
  year={2012},
  publisher={IOP Publishing}
}

@article{mason1995,
  title={Optical measurements of frequency-dependent linear viscoelastic moduli of complex fluids},
  author={Mason, Thomas G and Weitz, David A},
  journal={Physical review letters},
  volume={74},
  number={7},
  pages={1250},
  year={1995},
  publisher={APS}
}

@article{mason2000,
  title={Estimating the viscoelastic moduli of complex fluids using the generalized Stokes--Einstein equation},
  author={Mason, Thomas G},
  journal={Rheologica acta},
  volume={39},
  number={4},
  pages={371--378},
  year={2000},
  publisher={Springer}
}

@article{dasgupta2002,
  title={Microrheology of polyethylene oxide using diffusing wave spectroscopy and single scattering},
  author={Dasgupta, Bivash R and Tee, Shang-You and Crocker, John C and Frisken, BJ and Weitz, DA},
  journal={Physical review E},
  volume={65},
  number={5},
  pages={051505},
  year={2002},
  publisher={APS}
}

@article{helfer2025,
  title={Expanding the reach of diffusing wave spectroscopy and tracer bead microrheology},
  author={Helfer, Manuel and Zhang, Chi and Scheffold, Frank},
  journal={arXiv preprint arXiv:2502.14973},
  year={2025}
}

@article{tassieri2018,
  title={i-Rheo GT: Transforming from time to frequency domain without artifacts},
  author={Tassieri, Manlio and Ram{\'\i}rez, Jorge and Karayiannis, Nikos Ch and Sukumaran, Sathish K and Masubuchi, Yuichi},
  journal={Macromolecules},
  volume={51},
  number={14},
  pages={5055--5068},
  year={2018},
  publisher={ACS Publications}
}

@article{Scheffold2026,
  title = {Expanding the reach of diffusing wave spectroscopy and tracer bead microrheology},
  author = {Helfer, M. and Zhang, C. and Scheffold, F.},
  journal = {Phys. Rev. Res.},
  volume = {7},
  issue = {4},
  pages = {043274},
  numpages = {11},
  year = {2025},
  month = {Dec},
  publisher = {American Physical Society},
  doi = {10.1103/k34t-ghws},
  url = {https://link.aps.org/doi/10.1103/k34t-ghws}
}

@book{rubinstein2003,
  title={Polymer physics},
  author={Rubinstein, Michael and Colby, Ralph H},
  year={2003},
  publisher={Oxford university press}
}

@article{kapnistos2009,
  title={Nonlinear rheology of model comb polymers},
  author={Kapnistos, M and Kirkwood, KM and Ramirez, J and Vlassopoulos, D and Leal, LG},
  journal={Journal of rheology},
  volume={53},
  number={5},
  pages={1133--1153},
  year={2009},
  publisher={AIP Publishing}
}

@article{rubinstein1988,
  title={Self-consistent theory of polydisperse entangled polymers: Linear viscoelasticity of binary blends},
  author={Rubinstein, Michael and Colby, Ralph H},
  journal={The Journal of chemical physics},
  volume={89},
  number={8},
  pages={5291--5306},
  year={1988},
  publisher={American Institute of Physics}
}

@inbook{Debye,
url = {https://doi.org/10.4159/harvard.9780674366701.c48},
title = {Some Results of a Kinetic Theory of Insulators (Preliminary Communication)},
booktitle = {A Source Book in Chemistry, 1900-1950},
author = {P. Debye},
editor = {Henry M. Leicester},
publisher = {Harvard University Press},
address = {Cambridge, MA and London, England},
pages = {116--123},
doi = {doi:10.4159/harvard.9780674366701.c48},
isbn = {9780674366701},
year = {1968},
lastchecked = {2026-04-02}
}

@article{macdonald1990review,
  title={Review of mechanistic analysis by electrochemical impedance spectroscopy},
  author={Macdonald, Digby D},
  journal={Electrochimica Acta},
  volume={35},
  number={10},
  pages={1509--1525},
  year={1990},
  publisher={Elsevier}
}

@article{zhong2008optical,
  title={Optical Kerr effect spectroscopy of simple liquids},
  author={Zhong, Qin and Fourkas, John T},
  journal={The Journal of Physical Chemistry B},
  volume={112},
  number={49},
  pages={15529--15539},
  year={2008},
  publisher={ACS Publications}
}

@article{smith2021,
  title={i-RheoFT: Fourier transforming sampled functions without artefacts},
  author={Smith, Matthew G and Gibson, Graham M and Tassieri, Manlio},
  journal={Scientific Reports},
  volume={11},
  number={1},
  pages={24047},
  year={2021},
  publisher={Nature Publishing Group UK London}
}

@article{auhl2008linear,
  title={Linear and nonlinear shear flow behavior of monodisperse polyisoprene melts with a large range of molecular weights},
  author={Auhl, Dietmar and Ramirez, Jorge and Likhtman, Alexei E and Chambon, Pierre and Fernyhough, Christine},
  journal={Journal of Rheology},
  volume={52},
  number={3},
  pages={801--835},
  year={2008},
  publisher={AIP Publishing}
}

@article{baumgaertelDeterminationDiscreteRelaxation1989,
  title = {Determination of Discrete Relaxation and Retardation Time Spectra from Dynamic Mechanical Data},
  author = {Baumgaertel, M. and Winter, H. H.},
  year = 1989,
  month = nov,
  journal = {Rheologica Acta},
  volume = {28},
  number = {6},
  pages = {511--519},
  issn = {0035-4511, 1435-1528},
  doi = {10.1007/BF01332922},
  urldate = {2023-06-15}
}

@article{honerkampNonlinearRegularizationMethod1993,
  title = {A Nonlinear Regularization Method for the Calculation of Relaxation Spectra},
  author = {Honerkamp, J. and Weese, J.},
  year = 1993,
  month = jan,
  journal = {Rheologica Acta},
  volume = {32},
  number = {1},
  pages = {65--73},
  issn = {1435-1528},
  doi = {10.1007/BF00396678},
  urldate = {2025-06-27}
}

@article{kamathDeterminationPolymerRelaxation1989a,
  title = {The Determination of Polymer Relaxation Moduli and Memory Functions Using Integral Transforms},
  author = {Kamath, V. M. and Mackley, M. R.},
  year = 1989,
  month = jan,
  journal = {Journal of Non-Newtonian Fluid Mechanics},
  volume = {32},
  number = {2},
  pages = {119--144},
  issn = {0377-0257},
  doi = {10.1016/0377-0257(89)85032-3},
  urldate = {2026-04-07}
}

@article{shanbhagRelaxationSpectraUsing2020,
  title = {Relaxation Spectra Using Nonlinear {{Tikhonov}} Regularization with a {{Bayesian}} Criterion},
  author = {Shanbhag, Sachin},
  year = 2020,
  month = aug,
  journal = {Rheologica Acta},
  volume = {59},
  number = {8},
  pages = {509--520},
  issn = {1435-1528},
  doi = {10.1007/s00397-020-01212-w},
  urldate = {2025-06-27}
}

@article{takehComputerProgramExtract2013,
  title = {A {{Computer Program}} to {{Extract}} the {{Continuous}} and {{Discrete Relaxation Spectra}} from {{Dynamic Viscoelastic Measurements}}},
  author = {Takeh, Arsia and Shanbhag, Sachin},
  year = 2013,
  month = apr,
  journal = {Applied Rheology},
  volume = {23},
  number = {2},
  publisher = {De Gruyter Open Access},
  issn = {1617-8106},
  doi = {10.3933/applrheol-23-24628},
  urldate = {2025-06-02}
}

@CONTROL{REVTEX41Control}

@CONTROL{aip41Control,pages="1",title="0"}

\end{document}